# Uncovering the Electronic and Structural Origins of Oxygen-Vacancy Formation Energies through Data-Driven Learning

Zhenhao Zhou[1,2†], Yi Li[1,2†], Yang Cao[3], Zhenpeng Yao[1,2*]

[1]Center of Hydrogen Science and School of Materials Science and Engineering, Shanghai Jiao Tong University, Shanghai 200240, China

[2]Innovation Center for Future Materials, Zhangjiang Institute for Advanced Study, Shanghai Jiao Tong University, Shanghai 201203, China

[3]Acceleration Consortium, 700 University Avenue, Toronto, Ontario M7A 2S4, Canada

[†]These authors contributed equally to this work. * E-mail: yaozhenpeng@gmail.com.

## Abstract

Point defects play important roles in a wide range of materials applications. The rich electronic and structural behavior of oxygen vacancies makes them a useful case study for understanding point-defect energetics. Here, we systematically separated their formation energies into relaxation and non-relaxation contributions and examined the information needed to predict each. We established the first large-scale database of experimental oxide phases to combine configurational searches using bond distortions and atomic rattling, complete sampling of inequivalent oxygen sites, and supercells with defect-image separations exceeding 10 Å. The database contains 1,005 vacancy sites in 439 oxides. We identified distinct sets of descriptors for the two contributions. The models relate fixed-ion removal energies to pristine electronic spectra and eigenvalues, while data-driven analysis identifies band-edge positions as effective descriptors of the energy-level information. Structural relaxation emerges as the main bottleneck in predicting total formation energies. Using these descriptors, we trained multiple machine-learning models to predict the individual contributions and the total formation energy. Our models outperform defect graph neural networks and other models based on oxide formation enthalpy, oxygen p-band centres, band gaps and oxygen-cation electronegativity differences. We also achieve more accurate total formation energies than simulations using the machine-learning interatomic potentials (MLIPs) EqV3 for structural relaxation and energy evaluation. Finally, we provide relaxation trajectories recording atomic positions, energies, forces and stresses at every ionic step, offering the data for developing MLIPs in defective structures.

## Introduction

Point defects are central to the thermodynamic[1,2], electronic[3–5], ionic[6–9], and catalytic behavior[10,11] of crystalline materials. Their influence spans a wide range of concentrations and includes changes in electronic states, ionic transport[12], chemical reactivity[13], phase stability[14,15], and degradation[16,17]. In the dilute equilibrium limit, a defect's concentration depends exponentially on its formation free energy: lower formation free energy favors a higher equilibrium population at fixed temperature and site density[18–21]. Consequently, relatively small errors in the calculated energetics can translate into orders-of-magnitude differences in predicted defect concentrations and, consequently, material properties[18,22]. Accurate defect formation energies therefore provide a fundamental basis for understanding and predicting defect-controlled behavior in functional materials[22].

Reliable first-principles calculations of vacancy formation energies require attention to both the sites at which vacancies form and their relaxed configurations. If a crystal contains several symmetry-inequivalent sites, their vacancy formation energies may differ; enumerating all such sites is necessary to identify the energetically preferred vacancy and the full site-to-site variation within the parent structure[18,22,23]. The relaxed defect geometry can depend strongly on the initial configuration because standard local optimization algorithms explore only the basin of the potential-energy surface that is accessible from the starting structure[24–28]. A defect model generated directly from the pristine lattice may therefore remain trapped in a metastable local minimum and fail to recover a lower-energy reconstruction. Changes in local electronic occupation and bonding may drive dimerization, bond rearrangement, polaron localization, or Jahn-Teller-type distortions[6,24,29–31]. Such transformations may not emerge spontaneously during a conventional relaxation, and tighter convergence criteria alone do not resolve the underlying configurational sampling problem[22,24,32,33]. At fixed chemical potential and charge state, a higher-energy metastable geometry overestimates the ground-state formation energy and may change the ranking of inequivalent sites. Comprehensive site enumeration and systematic searches for low-energy reconstructions are therefore complementary requirements for reliable reference data. Methods such as ShakeNBreak[34] can deliberately perturb the local defect environment before relaxation, allowing symmetry breaking and alternative bonding motifs to be sampled more efficiently[22,33].

The need for comprehensive site sampling and configuration searching, however, further increases the already substantial computational cost of defect calculations. Compared with pristine bulk calculations, vacancy calculations often require large supercells[23,35], reduced symmetry[24,36], and independent structural optimizations for each inequivalent defective configuration[24]. To reduce the need for costly defect calculations, previous studies have developed machine-learning models and physical descriptors to predict defect formation energies from composition[37,38], pristine crystal structures[39–41], local chemical environments[37,42–53], or electronic-structure information[54–59]. Physically motivated regressions and graph-based models have achieved encouraging predictive performance[39,54]. Nevertheless, datasets can differ in the experimental provenance of parent structures, coverage of inequivalent vacancy sites, and extent of defect configurational searching. These differences, together with choices of features and model architecture, complicate model comparison and transferability. A value obtained for one selected site cannot necessarily represent either the lowest vacancy formation energy or its variation across an oxide. More fundamentally, a defect formation energy combines energetic contributions arising from distinct physical processes, including changes in electronic structure and bonding upon atom removal and the subsequent structural response of the defective lattice. Predicting only the total energy may obscure which information is needed to describe each contribution. We therefore ask why vacancy formation energies remain difficult to predict and which physical information is required for accurate and transferable predictions.

Oxides offer a broad experimental materials space, spanning diverse compositions, structural families, cation valences, coordination environments, and bonding motifs[60–63]. This diversity enables a stringent test of whether predictive relationships extend beyond narrowly defined material classes. Among defects in oxides, oxygen vacancies are especially relevant because they modify local electronic states[4,62,63], mediate oxide-ion transport[64], influence oxygen exchange and catalytic reactions[65,66], and contribute to oxygen-loss-related degradation[17]. Oxygen removal also produces site-dependent charge redistribution and structural relaxation[61,62]. Considering every symmetry-inequivalent oxygen site is thus important for finding the lowest vacancy formation energy of each parent oxide and capturing variations in local response. Oxygen vacancies therefore combine technological significance with the physical diversity needed to examine the information requirements of defect-energy prediction.

Previous studies[30,37] have separated vacancy formation energies into fixed-ion atom-removal and structural-relaxation contributions. For oxygen vacancies, the fixed-ion contribution ($E_s$) describes electronic and bonding changes upon oxygen removal at unchanged ionic positions and lattice parameters, whereas the relaxation contribution ($E_r$) is the energy change upon subsequent structural relaxation. Building on this decomposition, we examine which descriptors available before defect relaxation can predict each term and when structural information from relaxed defective structures becomes necessary.

Here, we establish a quality-controlled database of oxygen-vacancy formation energies for experimentally reported oxide structures from the Materials Project[67] (Fig. 1a). For each included experimental parent structure, our automated workflow enumerates all crystallographically inequivalent oxygen sites and explores both direct relaxation and multiple locally perturbed[24,34] vacancy configurations. The resulting dataset contains 439 oxide structures and 1,005 site-resolved vacancy models, supported by more than 15,000 first-principles calculations and accompanying electronic-structure and bonding analyses. Using these reference data, we investigate $E_s$ and $E_r$ separately. COHP, oxygen p-DOS, and band-edge positions relative to the Fermi level predict $E_s$ with $R^2$ = 0.935, whereas $E_r$ is harder to infer from information available before defect relaxation. With atomic forces included, direct predictions of total formation energy $E_f$ reach $R^2$ = 0.864 on held-out parent structures. These findings demonstrate the importance of complete inequivalent-site sampling and reliable structural relaxation, and identify defect-induced structural response as a central information bottleneck for predicting vacancy formation energies across chemically diverse oxides.

## Construction of a high-quality oxygen-vacancy formation energy database

To establish a reliable and chemically diverse dataset of oxygen-vacancy formation energies, we developed an automated high-throughput first-principles workflow integrating experimental-structure selection, symmetry-based site enumeration, supercell construction, configurational searching, electronic-structure analysis, and data curation (Fig. 1a). Experimentally reported oxide structures from the Materials Project[67] were randomly divided into computational batches. A Python workflow based on pymatgen[68] automated structure processing, vacancy-model construction, and the preparation of VASP[69] input files, including element-specific pseudopotentials, initial magnetic moments, and Hubbard U parameters where applicable. Automated task management collected calculation outputs, detected convergence failures, and organized tasks requiring correction or resubmission.

The structural scope of the database was defined by experimental provenance. Every parent oxide retained for analysis was associated with an experimentally reported crystal structure in the Materials Project. This selection covers a range of experimentally realized oxide compositions and crystal structures. It provides a consistent starting point for comparing vacancy energetics across distinct materials, rather than relying on hypothetical parent structures. To mitigate spurious interactions between periodically repeated defects, each pristine structure was expanded into a supercell for which the minimum separation between vacancy images in the initial defect model exceeded 10 Å (Fig. 1c). Figure 1c shows the fitted decrease in vacancy-site ranking deviation relative to the largest tested supercells as the cells are enlarged. This criterion refers to the initially generated supercells, as both the atomic positions and lattice parameters were subsequently allowed to relax. For each parent oxide, all crystallographically inequivalent oxygen sites were identified and indexed through symmetry analysis (Fig. 1b). Consistent with previous studies[37,39,70], we included only neutral oxygen vacancies in the dataset. A separate neutral oxygen-vacancy model was generated by removing one oxygen atom at each such site. By retaining every inequivalent site, the dataset resolves the full site-to-site variation in vacancy formation energies within each parent structure and enables identification of its energetically preferred vacancy site, rather than treating one arbitrarily selected site as representative of the entire oxide.

To address the sensitivity of defect energetics to the initial atomic configuration, we performed both conventional direct relaxation and a systematic search over locally perturbed vacancy configurations for every inequivalent oxygen site (Fig. 1d). The trial configurations were generated with the ShakeNBreak approach[24,34] and comprised the full set of structures produced by the adopted perturbation protocol, rather than an assumed exhaustive search of the entire configurational landscape. All trial structures were relaxed using DFT, with both ionic positions and lattice parameters optimized. Among the converged, physically valid configurations, the lowest-energy structure identified for each site was selected for subsequent analysis. The direct-relaxation results provided a reference for quantifying the influence of structural searching. As shown in Fig. 1e, lower-energy configurations were identified for 869 of the 1,005 vacancy sites (86.47%), and the energetic ordering of inequivalent oxygen sites changed in 83 (of 439) oxide structures. These comparisons demonstrate that site enumeration and configurational searching address complementary sources of uncertainty in the resulting vacancy-energy labels. Perturbed calculations accounted for 94.07% of the total relaxation runtime (Fig. 1g).

The workflow involved more than 15,000 first-principles calculation tasks, including pristine and defective structural optimizations and total-energy calculations. Density-of-states (DOS) calculations and crystal orbital Hamilton population (COHP) [71] analyses using LOBSTER[72] were performed for all pristine structures and the selected lowest-energy defect configurations, providing electronic-structure and chemical-bonding information for both states. Calculation convergence and structural integrity were systematically checked, while potentially anomalous results, including severe structural distortions or unphysical interatomic separations, were flagged for manual review. Physically plausible defect-induced reconstructions were distinguished from calculation artifacts, and only configurations confirmed to be invalid were excluded.

Following quality control, the curated database comprised 439 experimentally reported oxide structures and 1,005 site-resolved oxygen-vacancy formation-energy entries, each obtained by considering every crystallographically inequivalent oxygen site and selecting the lowest-energy valid configuration found in the sampled search, generating a comprehensive collection of structural, energetic, electronic, and bonding information associated with oxygen-vacancy formation. Together with the associated pristine/defective geometries, DOS and COHP information, these data provide a consistent basis for investigating the distinct electronic and structural-relaxation contributions to oxygen-vacancy formation energies.

## Results

The oxygen-vacancy formation energy is separated into fixed-ion removal and structural-relaxation contributions:

$$E_f = E_s + E_r, \tag{1}$$

$$E_s = \mathcal{E}_D^{\text{frozen}} - \mathcal{E}_P + \mu_O, \tag{2}$$

$$E_r = \mathcal{E}_D^{\text{relaxed}} - \mathcal{E}_D^{\text{frozen}}. \tag{3}$$

Here $\mathcal{E}_P$ is the total energy of the relaxed pristine structure. $\mathcal{E}_D^{\text{frozen}}$ and $\mathcal{E}_D^{\text{relaxed}}$ are the total energies of the frozen and relaxed defective structures, respectively. The frozen defective structure is formed by removing one oxygen atom while keeping the remaining atomic positions and lattice parameters unchanged. $\mu_O$ is the chemical potential of oxygen in the external reservoir.

Tight-binding models have been used to describe oxygen–metal coupling and the electronic changes associated with oxygen vacancies[73,74]. To examine what information is needed to describe the fixed-ion removal energy $E_s$, we first consider a minimal model with one oxygen $2p$ orbital $|O\rangle$ coupled to an effective host orbital $|H\rangle$ (Fig. 2a). For noninteracting electrons, the one-electron Hamiltonian is

$$H_P = \begin{pmatrix} a & t & 0 & 0 \\ t^* & b & 0 & 0 \\ 0 & 0 & a & t \\ 0 & 0 & t^* & b \end{pmatrix} = \begin{pmatrix} h_P & 0 \\ 0 & h_P \end{pmatrix}. \tag{4}$$

Here $h_P$ acts on $(|O\rangle, |H\rangle)$; $a$ and $b$ are their uncoupled energies, and $t$ is the coupling. The two identical spin blocks give each spatial eigenvalue two spin states:

$$h_P|n\rangle = \epsilon_n|n\rangle, \qquad n = \pm, \qquad \epsilon_- \le \epsilon_+. \tag{5}$$

In the ground state, two electrons occupy $\epsilon_-$, giving $E_P = 2\epsilon_-$. Removing the oxygen orbital and one electron, leaves $E_D = b$. The fixed-ion removal energy is therefore

$$E_s^{(2)} = E_D - E_P + \mu_O = b - 2\epsilon_- + \mu_O, \tag{6}$$

The electron counts refer to the orbitals retained in the model.

The uncoupled host energy $b$ is an input to this model. To connect the model with a pristine DFT calculation, we first use the trace relation

$$\text{Tr}h_P = a + b = \epsilon_- + \epsilon_+. \tag{7}$$

This gives $b = \epsilon_- + \epsilon_+ - a$. To obtain the remaining uncoupled energy $a$ from the same coupled system, we use the oxygen p-DOS, summed over both spins:

$$D_p(E) = 2\sum_{n=\pm} |\langle O|n\rangle|^2 \delta(E - \epsilon_n). \tag{8}$$

The spectral decomposition of $h_P$ gives

$$a = \langle O|h_P|O\rangle = \frac{1}{2}\int E\, D_p(E)\, dE. \tag{9}$$

All integrals include occupied and unoccupied states on the same energy reference. Substituting into the trace relation gives

$$b = \epsilon_- + \epsilon_+ - \frac{1}{2}\int E\, D_p(E)\, dE, \tag{10}$$

$$E_s^{(2)} = \mu_O + \epsilon_+ - \epsilon_- - \frac{1}{2}\int E\, D_p(E)\, dE. \tag{11}$$

For $t \neq 0$, both pristine levels appear in the oxygen p-DOS. Their energies and the DOS integral therefore determine $E_s^{(2)}$.

We next consider two oxygen states $|O_1\rangle, |O_2\rangle$ and two host states $|H_-\rangle, |H_+\rangle$ (Fig. 2b), with the same spin structure. Choosing the eigenstates of the separate oxygen and host blocks as the basis gives

$$h_P = \begin{pmatrix} \alpha_1 & 0 & t_{1-} & t_{1+} \\ 0 & \alpha_2 & t_{2-} & t_{2+} \\ t_{1-}^* & t_{2-}^* & b_- & 0 \\ t_{1+}^* & t_{2+}^* & 0 & b_+ \end{pmatrix} = \begin{pmatrix} A & T \\ T^\dagger & B \end{pmatrix}. \tag{12}$$

The diagonal blocks $A$ and $B$ contain the oxygen energies $\alpha_1, \alpha_2$ and host energies $b_- \leq b_+$, respectively. The matrix $T$ contains all four oxygen–host couplings. The spatial eigenstates satisfy $h_P|n\rangle = \epsilon_n|n\rangle$, with $\epsilon_1 \leq \epsilon_2 \leq \epsilon_3 \leq \epsilon_4$. Four electrons occupy the two lowest levels. Removing the two oxygen states and two electrons leaves the unchanged host block $h_D = B$, with both remaining electrons in $b_-$:

$$E_P = 2(\epsilon_1 + \epsilon_2), \qquad E_D = 2b_-, \qquad E_s^{(4)} = E_D - E_P + \mu_O. \tag{13}$$

To obtain the lower host level $b_-$, we need both the centre and separation of the two host levels (Fig. 2b). Define their sum and sum of squares as

$$S = \mathrm{Tr}\, B = b_- + b_+, \qquad Q = \mathrm{Tr}\,(B^2) = b_-^2 + b_+^2. \tag{14}$$

The centre is $S/2$, and the separation and remaining energy are

$$\varDelta_B = b_+ - b_- = \sqrt{2Q - S^2}, \qquad E_D = S - \varDelta_B. \tag{15}$$

Let $\varPi_O = \sum_{i=1}^{2} |\, O_i\rangle\langle O_i|$ project onto the oxygen states. Their summed p-DOS and the host DOS are

$$D_p(E) = 2\sum_{n=1}^{4} \langle n|\, \varPi_O |n\rangle \delta(E - \epsilon_n), \tag{16}$$

$$D_H(E) = 2\sum_{n=1}^{4}\delta(E-\epsilon_n) - D_p(E). \tag{17}$$

The host DOS gives $S$ directly, whereas its $E^2$-weighted integral also contains the coupling contribution. This follows from the host block of $h_p^2$, which is $B^2 + T^\dagger T$:

$$\frac{1}{2}\int E\, D_H(E)\, dE = S, \qquad \frac{1}{2}\int E^2\, D_H(E)\, dE = Q + \mathrm{Tr}\,(T^\dagger T). \tag{18}$$

To remove this extra term, we use the oxygen–host coupling operator and its COHP spectrum[71]:

$$V = \begin{pmatrix} 0 & T \\ T^\dagger & 0 \end{pmatrix}, \qquad C(E) = 2\sum_{n=1}^{4}\langle n|V|n\rangle \delta(E-\epsilon_n). \tag{19}$$

This signed COHP includes both spins and both hopping directions. Its first energy-weighted integral isolates the coupling:

$$\int EC(E)\, dE = 2\mathrm{Tr}\,(V\, h_P) = 4\mathrm{Tr}\,(T^\dagger T). \tag{20}$$

Subtracting this contribution and substituting the expression for $D_H$ gives

$$S = \sum_{n=1}^{4}\epsilon_n - \frac{1}{2}\int E\, D_p(E)\, dE, \tag{21}$$

$$Q = \sum_{n=1}^{4}\epsilon_n^2 - \frac{1}{2}\int E^2\, D_p(E)\, dE - \frac{1}{4}\int EC(E)\, dE. \tag{22}$$

The removal energy then follows as

$$E_s^{(4)} = S - \sqrt{2Q - S^2} - 2(\epsilon_1 + \epsilon_2) + \mu_O. \tag{23}$$

The DOS integrals and pristine eigenvalues determine the host-level sum; COHP separates the coupling contribution needed to recover the host-level separation. The trace identities for $\mathrm{Tr}\, B$ and $\mathrm{Tr}\,(B^2)$ hold for host blocks of any size. Both models replace uncoupled host energies with quantities obtained from the pristine spectrum. In real oxides, orbital energies and the couplings of remaining energy levels vary with the neighboring elements and atomic positions. We therefore examine whether additional features can summarize both the key information carried by the pristine eigenvalues and the main effects of these variations, together with pristine oxygen p-DOS and COHP, describe the $E_s$ in real oxides.

**Tab. 1** lists the features considered in the ablation experiment (Selected oxygen: the oxygen forming the defect). Their representations in the models are described in the **Tab. S1.** We encoded the COHP and oxygen p-DOS spectra using integrals with normalized Gaussian weights (Supplementary Note A). Like many prior researches[11,57], we preserved the unoccupied states in COHP and p-DOS spectrums. It consists with the tight-binding models, which shows that these states in the pristine structure carry information about the energies of the remaining host orbitals. Also, our follow-up experiments have found truncating at the Fermi level reduce the accuracy (Fig. S5).

**Table 1.** Feature families evaluated in the ablation experiments.

| **Category** | **Feature and representation** |
|---|---|
| Electronic structure | Oxygen–metal COHP spectra |
| Electronic structure | Selected oxygen p-DOS spectra |
| Electronic structure | Band-edge positions: $E_{\text{CBM}} - E_F$ and $E_{\text{VBM}} - E_F$ of the pristine structure |
| Local geometry | Distances from the selected oxygen to oxygen neighbors |
| Local geometry | Distances from the selected oxygen to non-oxygen neighbors |
| Local geometry | Selected-oxygen coordination number |
| Local geometry | Selected-oxygen voronoi-cell volume in the pristine structure |
| Local geometry | Metal–oxygen–metal angles |
| Chemical properties | Bond-valence oxidation assignments[68,75] and reduction capacity |
| Chemical properties | Covalent radii[76] of non-oxygen neighbors |

We compared COHP and oxygen p-DOS alone and with each additional feature family using TabDPT[77]. These comparisons used 401 defect sites from 178 pristine structures (40% of all parent structures). We used five-fold cross-validation within these 178 pristine structures, keeping all sites from each parent in the same fold. For each of three random seeds, we calculated MAE from the combined predictions across all five validation folds. We then averaged the three MAE values. Adding the band-edge positions, $E_{\text{CBM}} - E_F$ and $E_{\text{VBM}} - E_F$ of the pristine structure, reduced the MAE from 0.440 to 0.329 eV (Fig. 2c). We used COHP, oxygen p-DOS and band-edge positions as the selected features set. We then added each of the seven remaining feature families in turn. These additions produced only small changes in MAE (Fig. 2d). The full combination search is summarized in Table S2.

We next evaluated $E_s$ prediction with the selected features set using TabDPT, XGBoost[78], and RBF SVR[79] (**Fig. 3a,d**). We compared these models with the defect graph neural network (dGNN)[39], which uses graphs of the pristine crystal as the representations. We randomly divided the 439 parent structures into five approximately equal groups, keeping all defect sites from each parent together. Two groups (40%) supplied feature selection above. Each of the remaining three groups (60%) served as validation in turn, with training on the other four groups. For TabDPT, XGBoost, and dGNN[39], we repeated these three runs with three model seeds, giving nine fits per model. For each seed, we combined predictions from the three validation groups to calculate MAE and $R^2$. We report the mean of the three seed-specific scores. RBF SVR is deterministic, so we used three fits and calculated its metrics from the combined predictions across the same three validation groups. The results show that all three models with the selected set outperformed dGNN. TabDPT gave an MAE of 0.240 eV and an $R^2$ of 0.935. The SHAP analysis[80] identified oxygen p-DOS as the largest contributor in all three models (**Fig. 3b**).

We applied the same feature-selection procedure to $E_r$ and selected atomic forces, Voronoi-cell volume and neighboring covalent radii (Fig. S3a). These forces were calculated in the defective structure after oxygen removal, before ionic relaxation. Using this feature set, TabDPT predicted $E_r$ with an MAE of 0.413 eV, and XGBoost and RBF SVR gave comparable errors (Fig. 3e). All three models predicted $E_r$ less accurately than $E_s$, despite the smaller variance of $E_r$ (Fig. S3b). We then added descriptors of structural changes during relaxation (Table 2) to the same feature set (Fig. S3c). This improved $E_r$ prediction in all three models, with TabDPT's $R^2$ increasing from 0.493 to 0.708 (Fig. 3e).

**Table 2.** Post-relaxation features describing structural changes.

| **Category** | **Feature and representation** |
|---|---|
| Atomic displacements | Histograms of displacement magnitudes for oxygen and non-oxygen atoms, separately in the first two neighbor shells around the removed oxygen |
| Pair-distance changes | Histograms of signed metal–O distance changes for neighbors in the first shell |
| Force–displacement products | Histogram of $\mathbf{F}_i^0 \cdot \mathbf{u}_i$ over all retained atoms, where $\mathbf{F}_i^0$ is the force before relaxation and $\mathbf{u}_i$ is the displacement during relaxation |
| Cell deformation | Principal cell stretches when relaxing the defective structures |

Having examined the fixed-ion atom removal and relaxation contributions separately, we return to the total oxygen-vacancy formation energy, $E_f = E_s + E_r$. Adding atomic forces to the features selected for $E_s$ reduced TabDPT's MAE from 0.769 to 0.743 eV in the feature-selection subset. Adding Voronoi-cell volume and covalent radii provided little further improvement (Fig. S4). Using the features selected for $E_s$ together with atomic forces, TabDPT predicted $E_f$ with an MAE of 0.435 eV and $R^2$=0.864 (Fig. 3f). It outperformed both dGNN variants, the linear model of Deml et al. with published or refitted coefficients, and a random forest using the site-adapted Deml descriptors evaluated by Baldassarri et al[54]. All benchmark comparisons followed the same validation protocol, with the linear-model refitted using only the training data in each run. Our models achieved the highest $R^2$ and lowest MAE in most oxide classes, and the lowest MedAE in every class (Tables S3–S5). The arrow in the TabDPT panel of Fig. 3f marks an outlier for oxygen site 1 in $Ba_2UCuO_6$. The two inequivalent oxygen sites share a U–O–Cu bridging framework and neighboring Ba environments, but the Cu–O bond is longer at site 1 than at site 2, which get a smaller prediction error. Site 1 has a larger mean initial force magnitude on its four nearest Ba atoms (1.283 versus 0.874 eV/Å), yet a similar overall root-mean-square displacement (0.106 versus 0.107 Å), a smaller displacement of the neighboring Cu atom (0.015 versus 0.164 Å), and a smaller energy reduction upon relaxation (0.490 versus 0.780 eV). The model may not have fully learned the distinct response of site 1 to its initial forces, leading it to overestimate the energy lowering upon relaxation.

Like this outlier, despite the accurate prediction of $E_s$, predicting $E_f$ remained more difficult. We then compared direct $E_f$ prediction with the sum of separately predicted $E_s$ and $E_r$, using COHP, oxygen p-DOS, band-edge positions and atomic forces as the common inputs for all three targets. The summed predictions gave comparable accuracy for TabDPT and lower errors for XGBoost and RBF SVR in predicting $E_f$ (Fig. 3c). Together with the larger prediction error for $E_r$, this points to relaxation-energy prediction as a major bottleneck in improving the accuracy of $E_f$.

## Discussion

The results show that the difficulty in predicting $E_f$ lies in predicting $E_r$. This raises the question of whether we can predict structural changes before performing relaxation calculations. We first examine this difficulty through specific examples. In Fig. 4(a), $Pr_3TaO_7$ and $Pr_3NbO_7$ share the same *Cmcm* structural prototype, with Ta and Nb occupying corresponding cation sites. The oxygen atoms selected for removal occupy corresponding 8*g* Wyckoff sites and are each coordinated to four Pr atoms. After oxygen removal and relaxation, $Pr_3TaO_7$ retains the original vacancy. However, in the relaxed $Pr_3NbO_7$ structure, three oxygen atoms have rearranged in a chain: one occupies the original vacancy, a second occupies the first atom's original site, and a third occupies the second atom's original site, leaving a vacancy at the other end of the chain. The resulting $E_r$ values differ by 1.524 eV. Figure 4(b) compares removal of two inequivalent oxygen atoms, $O_1$ and $O_2$, from the same pristine $Ba_3FeO_5$ structure (*Pnma*). Both sites have one Fe and five Ba neighbors within 3.1 Å, with Fe–O distances of 1.846 and 1.793 Å, respectively. Removing either oxygen reduces the coordination number of its neighboring Fe from four to three. After relaxation, the Fe remains threefold coordinated for $O_1$ removal, whereas for $O_2$ removal it gains a new oxygen neighbor and its coordination number returns to four. The separation between this Fe and its new oxygen neighbor decreases from 3.652 to 1.912 Å. The $E_r$ values therefore differ by 0.944 eV between these two defective structures. These examples show that atomic rearrangements following oxygen removal are highly sensitive to the initial environment, and that similar local coordination can lead to entirely different relaxed structures.

As expected, the difficulty in predicting $E_r$ associated with the structural responses discussed above also affects the prediction of $E_f$. Figure 5 presents three cases in which relaxation plays different roles in the total defect formation energy. In $Ba_2ZrO_4$, $E_f$ is almost entirely determined by $E_s$, which is $6.626$ eV, whereas $E_r$ is only $-0.080$ eV. Removing the bridging oxygen converts two $ZrO_6$ coordination units into $ZrO_5$ units. Both Zr retain their five remaining oxygen neighbors after relaxation, and the mean remaining Zr–O distance decreases only from approximately 2.111 to 2.099 Å. The frozen and relaxed structures are therefore very similar [Fig. 5(a)]. By contrast, the fixed-ion removal energy of $PbO_2$ is only 1.223 eV, but relaxation releases 3.566 eV, reducing the final $E_f$ to $-2.343$ eV [Fig. 5(b)]. This process involves extensive atomic rearrangements and changes in the coordination environments of Pb atoms. The Pb atoms replace their oxygen neighbors despite retaining the same coordination number. $SrPuO_4$ illustrates a case in which $E_s$ and $E_r$ have large magnitudes but nearly cancel, giving a small final $E_f = -0.713$ eV [Fig. 5(c)]. During this process, all cation displacements are no greater than 0.143 Å, whereas the maximum oxygen displacement reaches 2.62 Å. It moves into the original vacancy, shifting the vacancy to its previous position. Thus, a small total formation energy does not imply small energy changes in both the removal and relaxation stages, nor does it imply minor structural changes. These cases illustrate the diversity of relaxation contributions and the accompanying structural responses across different systems, and show that relaxation generally cannot be treated as a small correction to the fixed-ion removal energy. Examining $E_s$ and $E_r$ separately reveals differences that are difficult to discern from the magnitude of the total defect formation energy alone.

Obtaining reliable relaxed defective structures is therefore central to accurate $E_f$ calculations, but searching for these structures requires many DFT evaluations. Machine-learning interatomic potentials (MLIPs) offer a way to accelerate this search[25,81]. We explored this possibility using EquiformerV3 (EqV3)[82,83], which was pretrained on OMat24[84] and subsequently fine-tuned on MPtrj[85] and subsampled Alexandria[84,86]. We used this model without further fine-tuning to relax pristine and defective structures , we selected the lowest-energy perturbed configuration for calculating the $E_r$ and $E_f$ (Supplementary

Information). Although for the total formation energies, direct TabDPT prediction achieved a lower MAE than EqV3 calculations (0.435 versus 0.515 eV), EqV3 predicted a comparable $E_r$ with an MAE of 0.454 eV [Fig. 6(a)], to the 0.413 eV obtained by TabDPT [Fig. 3(e)]. The remaining errors may partly reflect limited coverage of defective configurations sampled during relaxation in its training datasets, as these configurations can differ substantially from stable structures. As a practical workaround, we trained TabDPT to predict the residual $E_f^{\mathrm{DFT}} - E_f^{\mathrm{ML}}$ using COHP, oxygen p-DOS, band-edge positions and the raw EqV3 formation energy, allowing EqV3's predictions to be corrected without retraining EqV3. Adding the predicted residual to $E_f^{\mathrm{ML}}$ gave an MAE of 0.282 eV, compared with 0.448 eV after a constant offset correction [Fig. 6(b)]. This improvement demonstrates that our selected site-specific features above contain information useful for correcting EqV3's formation energy predictions. However, this correction improves the final $E_f$ predictions to some degree, while the EqV3 energies and forces that drive structural relaxation remain unchanged. A more direct way to improve the description of defect relaxation would be training MLIPs on DFT energies and forces for defective structures sampled along relaxation trajectories. Our dataset providing these quantities in trajectories could therefore support advanced MLIPs that enable faster and more accurate calculations of oxygen-vacancy formation energies.

## Conclusion

We demonstrate that compact, physically motivated input sets outperform dGNN, the Deml linear model and a random forest using site-adapted Deml descriptors in predicting oxygen-vacancy formation energies.. With TabDPT, we also achieve higher accuracy than direct EqV3 calculations. Beyond improved performance, we explain why further progress remains difficult. We can capture the fixed-ion contribution accurately with electronic descriptors, but predicting relaxation requires anticipating structural rearrangements that can differ markedly even between similar initial environments. We thus distinguish the comparatively predictable electronic contribution from the structural response that limits total formation-energy prediction. This distinction directs future efforts towards better descriptions of defect relaxation, rather than treating formation energy as a single prediction target. We provide an experimental-oxide database that resolves inequivalent vacancy sites and competing relaxed configurations as a common reference for this work. Its relaxation trajectories support addressing the central challenge of structural response for the MLIPs researchers.

## Code availability

Code for this work is available at https://github.com/zhenpengyao/oxygen-vacancy-learning .

## Acknowledgements

Z.Y., Z.Z. and Y.L. were supported by the National Natural Science Foundation of China (grant number 52373228).

## Contributions

Z.Y. conceived the overall project and supervised the research. Z.Z. performed the modeling, machine learning experiments, and conducted comparative benchmarking. Y.L. prepared the vacancy formation energies calculation workflow and the database. Z.Y., Z.Z., and Y.L. contributed to the interpretation of experimental results. All authors contributed to the preparation of the manuscript.

## Competing interests

The authors declare no competing interests.

## Figures

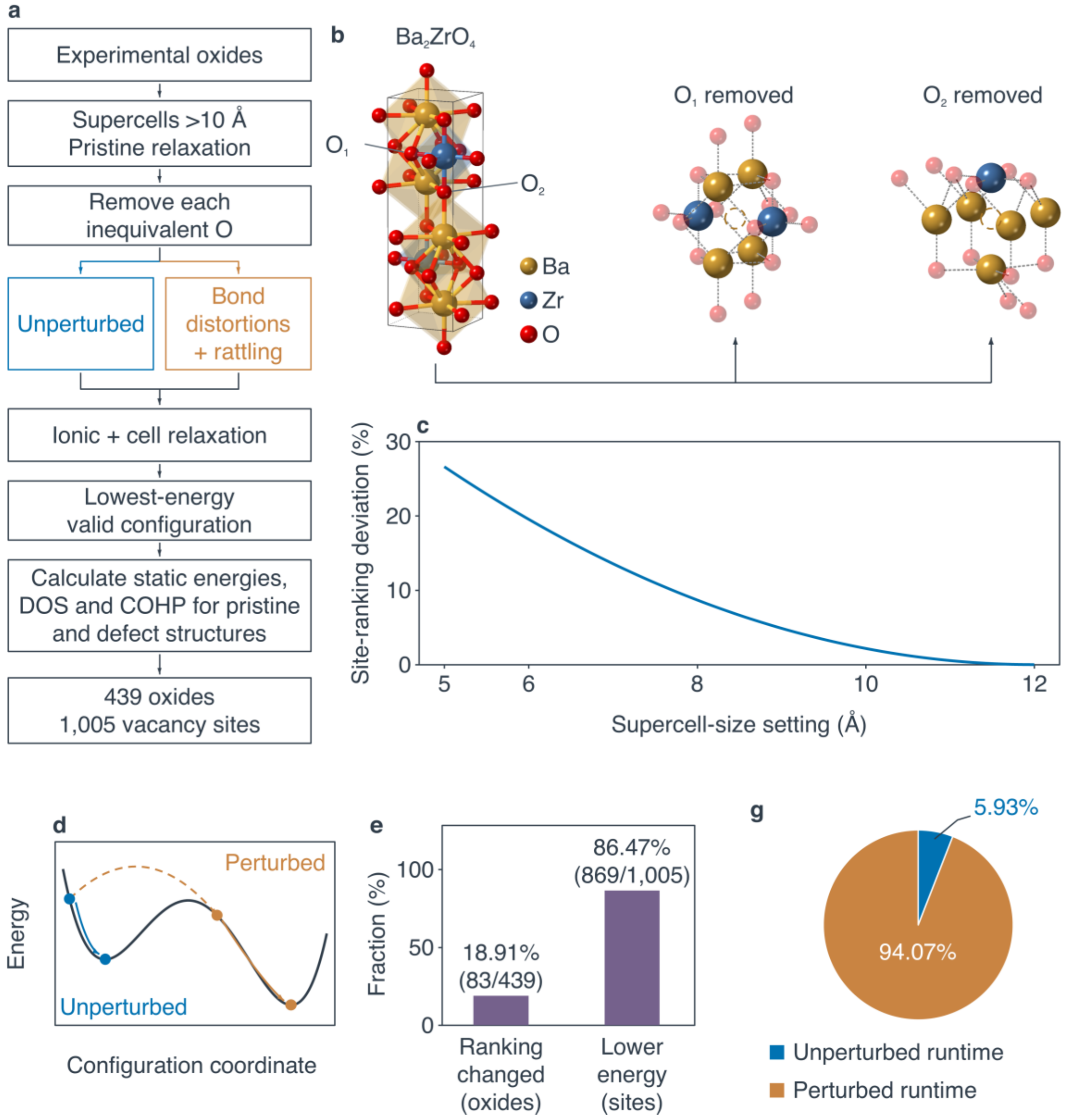

**Figure 1. Automated oxygen-vacancy database construction.** (a) Workflow. (b) Two inequivalent O sites in $Ba_2ZrO_4$ and the corresponding frozen defects. (c) Fitted supercell-size dependence of site-ranking deviation in $CaGa_4O_7$, $Ca_2Al_2O_5$, $Ba_2CoO_4$ and $Rb_2W_2O_7$. Deviation is the fraction of site pairs whose energy ordering differs from the largest-supercell reference. (d) Configurational-search schematic[24]. (e) Fractions of oxides with altered site rankings and vacancy sites with lower energies following the search. (g) Shares of relaxation runtime for unperturbed and perturbed calculations.

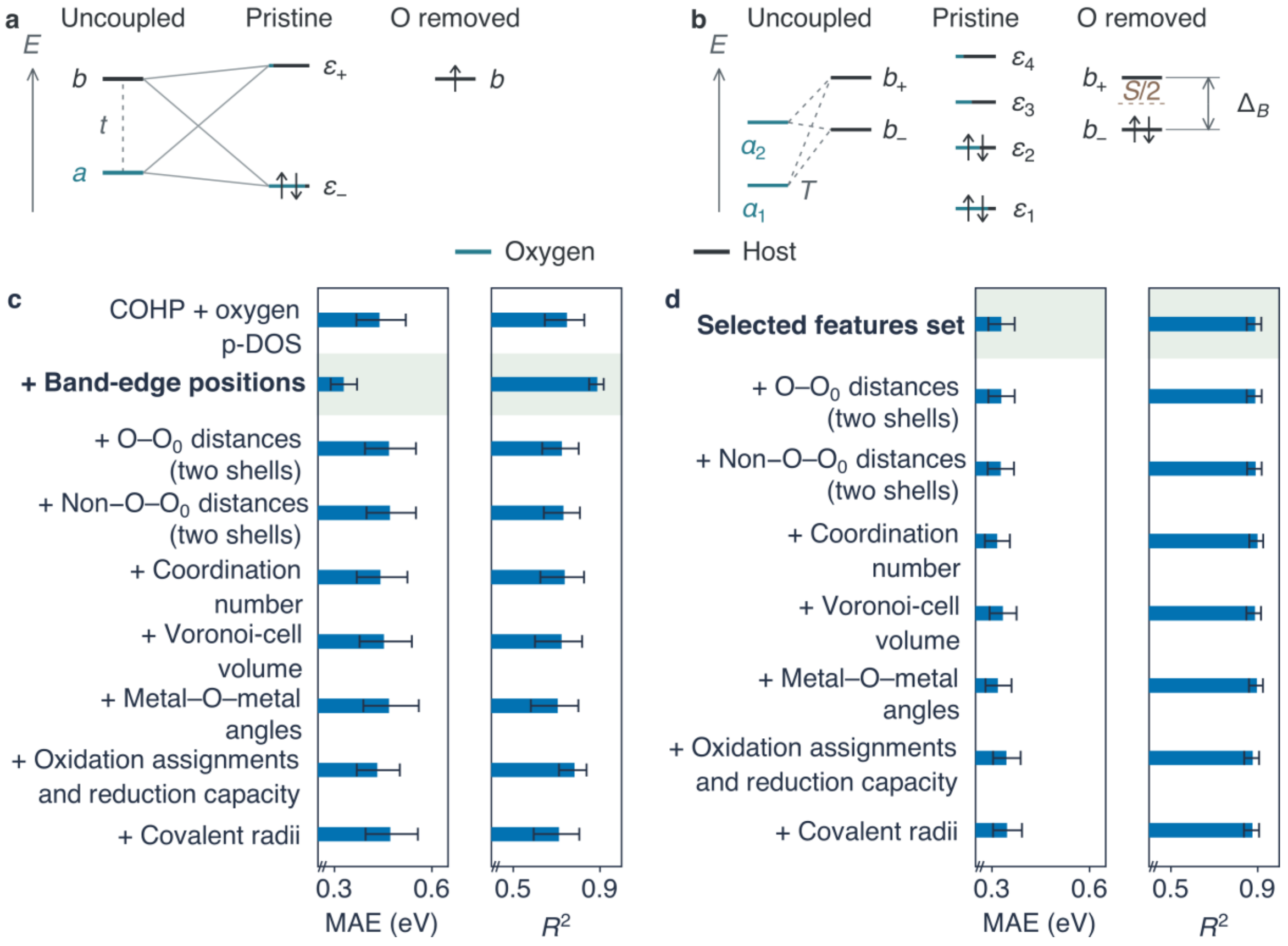

**Figure 2. Tight-binding models and feature selection for $E_s$.** (a) One oxygen orbital coupled to one host orbital. Dashed lines indicate the couplings, teal and charcoal segments show oxygen and host probabilities. Vertical arrows indicate electron spins. Oxygen removal leaves the host energies unchanged. (b) Two oxygen orbitals coupled to two host orbitals. $S/2$ and $\Delta_B$ denote the centre and separation of the remaining host levels. (c) TabDPT predictions of $E_s$ using COHP and oxygen p-DOS alone or with one additional pristine-derived feature family. (d) Predictions after adding each further pristine-derived feature family to the selected set: COHP, oxygen p-DOS and band-edge positions. Band-edge positions are $E_{\mathrm{CBM}} - E_{\mathrm{F}}$ and $E_{\mathrm{VBM}} - E_{\mathrm{F}}$ of the pristine structure. Both distance families include the first two neighbor shells. $O_0$ denotes the oxygen atom selected for removal. Whiskers show 95% confidence intervals.

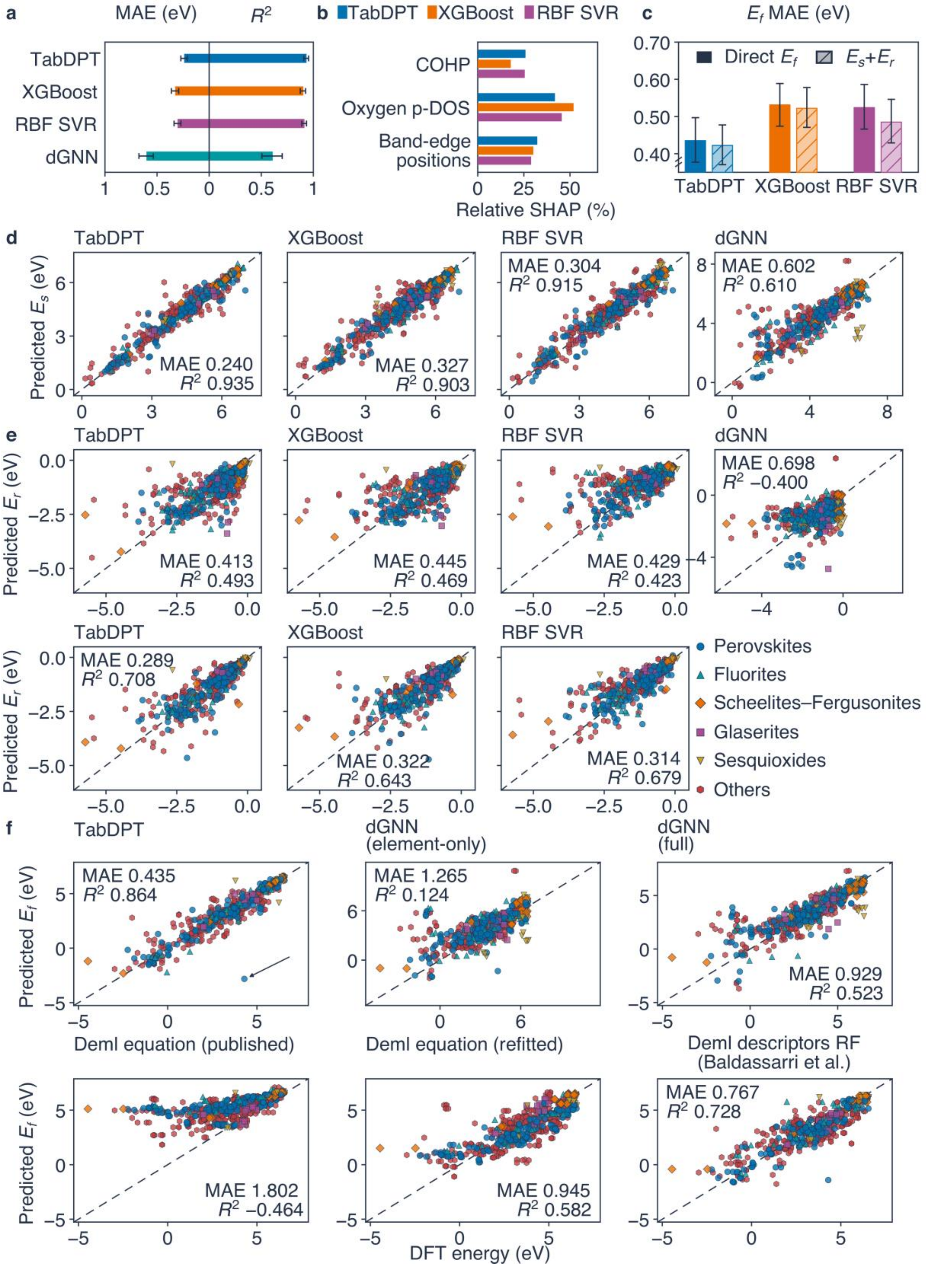

**Figure 3. Metrics of model performance in Fixed-ion removal $E_s$, relaxation $E_r$ and oxygen-vacancy formation $E_f$ energies**. (a) $E_s$ MAE and $R^2$. (b) SHAP contributions. (c) Direct $E_f$ versus $E_s + E_r$

(hatched), with all three targets predicted using the same features: COHP, oxygen p-DOS, band-edge positions and forces. (d) $E_s$ parity plots. (e) $E_r$before (upper) and after (lower) adding structural-change descriptors (**Table 2**). (f) $E_f$ benchmarks. TabDPT uses the inputs in (c). dGNN uses element-only encoding; full features with oxidation states, band gap, formation enthalpy, and a DOS-derived electron-mass descriptor. The Deml linear model uses published or refitted coefficients. Deml descriptors RF follows the site-specific adaptation evaluated by Baldassarri et al. Points metrics are averaged across seeds. Symbols denote oxide families, dashed lines indicate perfect agreement. Whiskers show 95% confidence intervals.

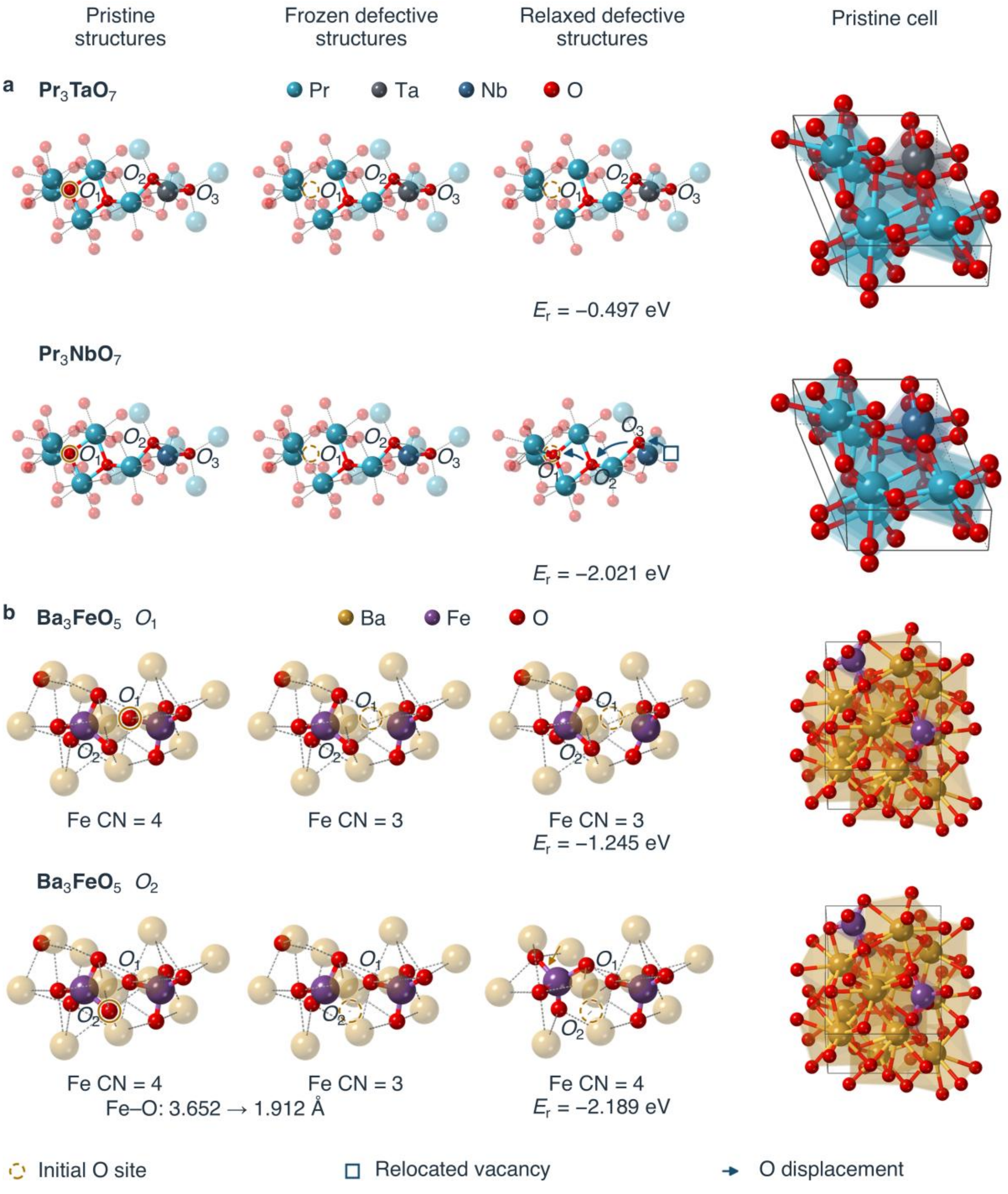

**Figure 4. Relaxation response examples.** (a) Corresponding oxygen sites in isostructural $Pr_3TaO_7$ and $Pr_3NbO_7$. (b) Inequivalent O1 and O2 sites in the same $Ba_3FeO_5$ parent. The first three columns show pristine structures, frozen defective structures, and relaxed defective structures. $\boldsymbol{E_r}$ is the relaxation energy, and CN is the coordination number of the Fe adjacent to the removed oxygen. The rightmost column shows pristine primitive cells.

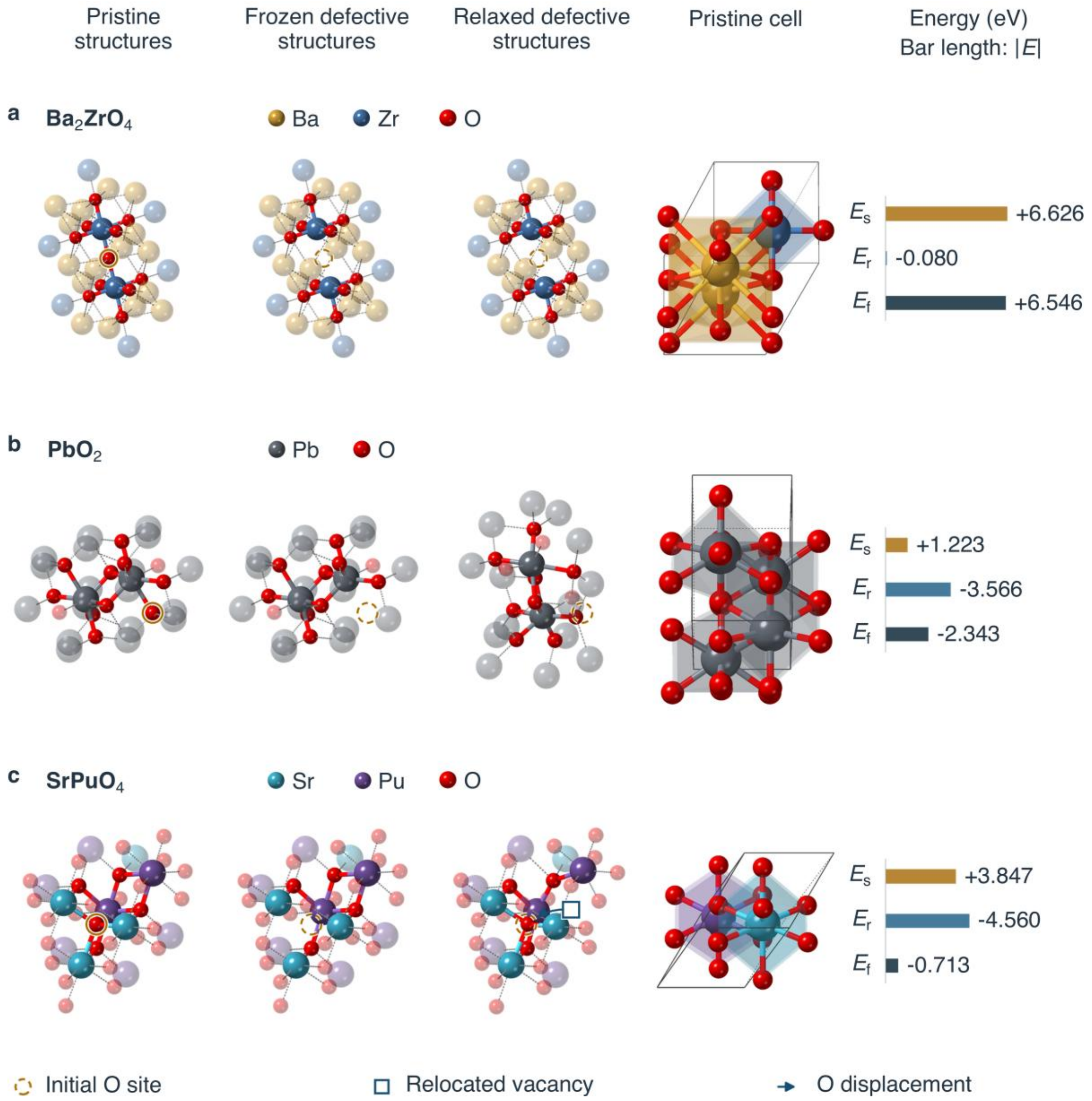

**Figure 5. Distinct contributions of structural relaxation to oxygen-vacancy formation energies.** Pristine structures, frozen defective structures, and relaxed defective structures are compared for (a) $Ba_2ZrO_4$, with minimal relaxation; (b) $PbO_2$, with extensive bond rearrangement; and (c) $SrPuO_4$, with oxygen relocation and near cancellation of the removal and relaxation energies. The fourth column shows pristine primitive cells. The rightmost column shows $E_s$, $E_r$ and $E_f = E_s + E_r$ of this defect.

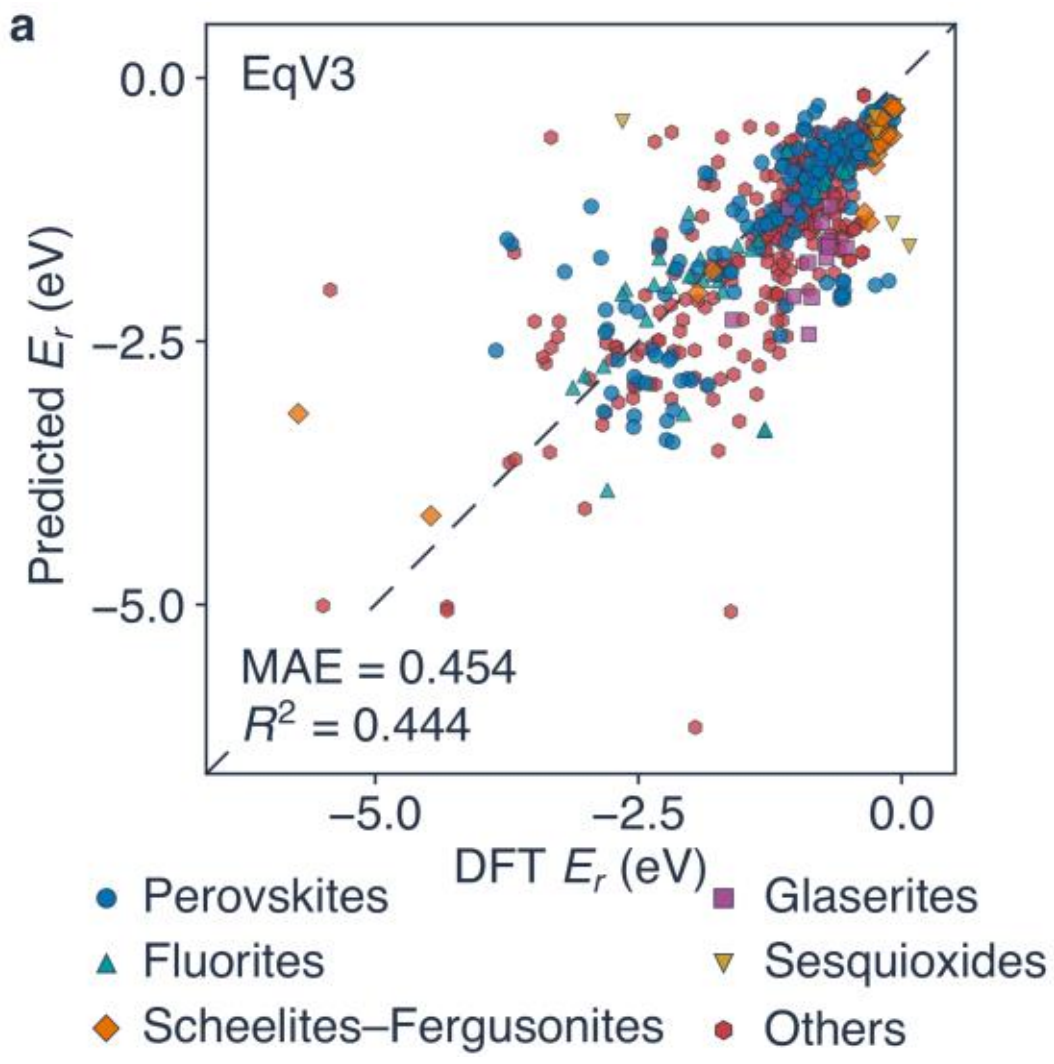

**Figure 6. (a) EqV3-predicted $E_r$ versus the DFT reference. (b) MAEs of $E_f$ predicted by EqV3, after a constant offset correction, and after adding the DFT–EqV3 residual predicted by TabDPT or XGBoost.** Both correction models use the selected set of COHP, oxygen p-DOS and pristine band-edge positions, together with the raw EqV3 formation energy. Corrections are fitted on training structures and evaluated on the same held-out structures. Symbols indicate oxide families. MAEs are in eV. Whiskers show 95% confidence intervals.